  \def\L{\Lambda}
  \def\om{\omega}

 \def\P{\Psi} 

\def\imo{i}

\def\be{\begin{equation}}
\def\ee{\end{equation}}
\def\bea{\begin{eqnarray}}
\def\eea{\end{eqnarray}}

\documentclass[11pt]{article}
\usepackage[dvips]{graphicx}
\usepackage{longtable}
\usepackage{amssymb}
\begin{document}
\hyphenation{Schwar-zschild}

\centerline{\textbf{\Large Scalar field perturbations  of the Schwarzschild }}
\centerline{\textbf{\Large black hole in the G\"{o}del Universe}}

\vspace{5mm}

\centerline{\textbf{R.A. Konoplya and Elcio Abdalla}}
\centerline{Instituto de Fisica, Universidade de S\~{a}o Paulo,} 
\centerline{ C.P.66318, CEP 05315, S\~{a}o Paulo SP, Brazil}
\centerline{konoplya@fma.if.usp.br}
\vspace{5mm}

\begin{abstract}
We investigate the scalar field perturbations of the $4+1$-dimensional
Schwarzschild black hole immersed in a G\"{o}del Universe, described by
the Gimon-Hashimoto solution.
This may model the influence of the possible rotation of the Universe 
upon the radiative processes near a black hole.
In the regime when the scale parameter $j$ of the G\"{o}del background
is small, the oscillation frequency is linearly decreasing with $j$, while
the damping time is increasing. The quasinormal modes are damping,
implying stability of the Schwarzschild-G\"{o}del space-time against
scalar field perturbations.  The approximate analytical formula
for large multipole numbers is found.   
\end{abstract}

\thispagestyle{empty}

\newpage

\section{Introduction}

Black hole's  behavior is often crucially dependent upon the cosmological
background in which the black hole is immersed.
The simple case of a black hole immersed in an asymptotically flat 
space-time is described by the Schwarzschild solution. A natural
extension is to consider a cosmological term which is 
described in terms of the Schwarzschild-de Sitter or Schwarzschild - anti-de
Sitter solutions. Recent investigations of radiative processes around
such black holes show that the radiative features are dependent 
upon the asymptotic conditions on infinity. Thus, for example, 
the quasinormal spectrum and late time tails are totally different for
Schwarzschild \cite{S}, Schwarzschild-de Sitter \cite{SdS}  and Schwarzschild - 
anti-de Sitter black holes \cite{SAdS}.

The cosmological background we are interested in here, is the rotating
Universe. The rotation seems to be a universal phenomenon: all
compact objects in the Universe rotate. Yet the standard 
Friedman-Robertson-Walker metric represents rather idealized model
of isotropic homogeneous world filled with perfect fluid. It 
looks improbable that such a finely tuned universe can exist
starting from the Big Bang to the present stage. In the beginning 
of the investigations of rotating cosmological models it was 
suggested that one should observe the anisotropy of  
the Microwave Background Radiation (MBR), yet, as 
shown later, that the rotating models with no anisotropy of MBR or
broken causality can exist \cite{obukhov1}. In addition, apparent anisotropy
in distribution of the observed angles between the polarization
vectors and position of the major axis of radio sources can be related
to a possible rotation of the Universe \cite{Brich}. For further advance in 
the possibility of observation of global rotation see the review \cite{obukhov2}.

An exact solution for the rotating Universe was found by G\"{o}del \cite{Godel}.
His solution was originally proposed for a four dimensional space-time. 
It possesses, among others, the following properties: it is 
homogeneous, has rotational symmetry, and allows the definition of the
direction of positive time consistently in the whole solution, and, 
what was in the focus of further research, it allows closed time-like
curves, i.e. the time machine.

Recently, the G\"{o}del Universes have been of considerable interest \cite{kucha1},
because in five-dimensional minimal supergravity the maximally supersymmetric
G\"{o}del-type universes are U-dual to pp-waves. Thereby, the G\"{o}del-type
universes are important as a an opportunity of quantizing strings in this
background a well as due to its relation to the corresponding limit of 
super-Yang Mills theories. On the gravitational side, the pp-waves 
dual to the G\"{o}del Universe corresponds to the Penrose limit 
of near-horizon geometries.

As far as we are aware, until recently, an exact 
solution for a stationary black hole immersed in a rotating 
Universe was not known. Nevertheless, such a solution has been obtained by Gimon and 
Hashimoto within the above mentioned five-dimensional minimal supergravity \cite{gimon}. 
This solution represents the $4+1$-dimensional Schwarzschild space-time 
when the scale parameter of the G\"{o}del background $j$ goes to zero,
and  to the five dimensional G\"{o}del Universe when the black hole mass 
vanishes, thereby giving us the model for the Schwarzschild black hole
immersed in the rotating Universe. The different features of this 
solution have been investigated recently in a series of papers \cite{kucha2}. 
The generalizations of the Gimon-Hashimoto solution were obtained in 
\cite{kucha3}.

Yet, here, we are interested in this solution from a rather different
point of view, namely, we would like to find out 
what will happen with black hole (classical) radiation 
in  the rotating Universe. 
The straightforward way to know it, is to investigate
the quasinormal modes which govern the black hole response to external 
perturbations at late times. The quasinormal spectrum is sensitive to boundary
conditions both at the event horizon and at spatial infinity, so
the spectrum must be considerably affected by the rotating
cosmological background.

In the present paper we had to be limited by the case of ``slow rotation'', i.e. 
the case when the influence of the cosmological background is weak.
This happens when the above mentioned parameter $j$ is small.
We found that at least in the regime of small $j$, the black hole 
is stable against scalar field perturbations and as a result all 
found modes are damping. Due to cosmological rotation the real
oscillation frequencies are decreasing and are roughly proportional to
$j$, while the damping rates are decreasing non-linearly with $j$.
In addition we derive an approximate analytical formula for QN modes
with large multipole number $L$. Fortunately the regime of small 
$j$ seems to be the most reasonable phenomenologically.

The paper is organized as follows: in Sec.2 we give some 
introductory material on the  Schwarzschild-G\"{o}del metric. 
In Sec 3., the scalar field equations is obtained in the limit
of small scale parameter $j$ of the cosmological background.   
For any $j$, the Klein-Gordon equation is not separable 
at least in the coordinates we have considered. In Sec.IV we find 
the quasinormal frequencies 
for the scalar field perturbations. In the end  we discuss the 
obtained results and future perspectives.

\section{Preliminaries of the Schwarzschild-G\"{o}del space-time.}

The bosonic fields of the minimal (4+1)- supergravity theory consist
of the metric and the one-form gauge field, which are governed by the
equations of motion
\begin{equation}
R_{\mu \nu} =2 \left(F_{\mu \alpha} F_{\nu}^{\alpha} -\frac{1} {6} g_{\mu \nu}
F^{2}\right)
\end{equation}
\begin{equation}
D_{\mu} F^{\mu \nu} =  \frac{1}{2 \sqrt{3}} \varepsilon^{\alpha \lambda
\gamma \mu \nu}  F_{\alpha \lambda} F_{\gamma \mu}  
\end{equation}
Here, $ \varepsilon_{\alpha \lambda \gamma \mu \nu} = \sqrt{-g} \epsilon_{\alpha \lambda \gamma \mu \nu}$.

In the Euler coordinates $(t, \theta', \psi', \phi')$, the solution of the
equations of motion (1), (2), describing the G\"{o}del universe, has the form \cite{gimon}: 
\begin{equation}
ds^2 = - (dt + j (r^2) \sigma_{L}^{3})^2 + dr^{2} + \frac{r^2}{4}
(d \theta'^{2} + d \psi'^{2} + d \phi'^{2} + 2 cos \theta' d \psi' d \phi'),
\end{equation}
where $\sigma_{L}^{3}= d \phi' + cos \theta d \psi'$. The parameter $j$ 
defines the scale of the G\"{o}del background.  At $j=0$ we have
the Minkowski space-time. The solution for the  Schwarzschild black hole 
in the G\"{o}del universe is given by \cite{gimon}
$$ ds^2 = - f(r) dt^2 -g(r) r \sigma_{L}^{3} d t + h(r) r^2
(\sigma_{L}^{3})^2 + k(r) d r^2 +$$ 
\begin{equation}
\frac{r^2}{4}
(d \theta'^{2} + d \psi'^{2} + d \phi'^{2} + 2 cos \theta' d \psi' d \phi'),
\end{equation}  
where 
$$f(r)=1- \frac{2 M}{r^2}, \quad  g(r) = 2 j r,$$
\begin{equation}
h(r) =j^2 (r^2 + 2 M),  \quad k(r) = \left(1 - \frac{2 M}{r^2} + \frac{16 j^2 M^2}{r^2}\right)^{-1}.  
\end{equation}     
The radius of the event horizon is also corrected by parameter $j$, 
\begin{equation}
r_{BH} = \sqrt{2 M (1- 8 j^2 M)}.  
\end{equation}
Note that the maximal value of the black hole mass $M$ is $1/8
j^2$. For a larger mass the horizon area vanishes and one has a naked 
singularity. 

The above black hole metric keeps five of the nine isometries of the G\"{o}del
universe, generated by $\partial_{t}$, and by four generators of the
$SU(2) \times U(1)$ subgroup of the $SO(4)$ isometry group acting on
$S^3$ \cite{gimon}. 
 
In the limit $j =0$ we have the (4+1)-dimensional Schwarzschild
solution, while in the
limit of $m=0$ the pure G\"{o}del space-time is recovered.
To treat the scalar field perturbations around such
a Schwarzschild-G\"{o}del black hole let us rewrite the metric in the 
bi-spherical coordinates ($\phi$, $\psi$, $\theta$), which are connected
with the Euler angles ($\phi'$, $\psi'$, $\theta'$) by
\begin{equation}  
\phi' = \psi + \phi, \quad \psi'= \psi - \phi, \quad, \theta'= 2
\theta.
\end{equation}    
Then, in the regime of small $j$,  i.e.discarding terms of order
$O(j^2)$, the metric takes the form
$$ ds^2 \approx - f(r) dt^2 -2 g(r) r ((sin \theta)^2 d \phi + (cos
\theta)^2 d \psi) d t + k(r) d r^2 +$$ 
\begin{equation}
r^2 (d \theta^{2} + (cos \theta)^2 d \psi^{2} + (sin \theta)^2 d \phi^{2}).
\end{equation}   
Note that in the above equation $k(r) = \left(1 - \frac{2
M}{r^2}\right)^{-1}$.

Up to $O(j^2)$, the inverse metric $g^{\mu \nu}$ has components
\begin{equation}
g^{11}=-\left(1 - \frac{2 M}{r^2} \right)^{-1} \quad g^{22}=1 - \frac{2 M}{r^2}
\quad g^{33} = \frac{1}{r^2}
\end{equation}
\begin{equation}
g^{44}=r^{-2} (cos \theta)^{-2} , \quad g^{55}=r^{-2} (sin \theta)^{-2},
\quad g^{14}=g^{15}=- 2 j \left(1- \frac{2 M}{r^2} \right)^{-1}. 
\end{equation}

\section{Scalar field perturbations of the Schwarzschild - G\"{o}del space-time.}

The scalar field perturbations in a curved background are governed by
the Klein-Gordon equation
\begin{equation}
\Box \Phi  \equiv \frac{1}{\sqrt{-g}} \left(g^{ \mu \nu} \sqrt{-g} 
\Phi,_{\mu}\right),_{ \nu}  = 0.
\end{equation}
Since the background metric has the Killing vectors  $\partial_{t}$,
$\partial_{\psi}$,  $\partial_{\phi}$, the wave function $\Psi$
can be represented in the form
\begin{equation}
\Phi \sim e^{i \omega t + i k \psi + i m \phi} Y(\theta) R(r). 
\end{equation}

Unfortunately, variables in the Klein-Gordon equation are not separable, at least in
the considered coordinates for the full Gimon-Hashimoto  metric. 
The separability is connected with the existence of the Killing tensor
\cite{bagrov}, and, it is possible that one can separate variables in some other 
privileged coordinate systems. Here we were limited to small values of
$j$, for which the variables in the Klein-Gordon equation can be
separated even in ordinary bi-spherical or Euler coordinates.
      
Using the expressions for metric coefficients (8-10), 
the scalar field equation (11) takes the form:
\begin{equation}
r^{-3} \frac{\partial }{\partial r} \left(\left(1-\frac{2 M}{r^2}\right)
r^{3} \frac{\partial R(r)}{\partial r} \right) + (\omega^2 + 4 j
\omega (k + m)) f^{-1}(r) R(r) + \frac{\lambda}{r^2} R(r) =0,     
\end{equation}
where the separation constant comes from the equation for angular
variables,
\begin{equation}
\frac{1}{cos \theta sin \theta} \frac{\partial}{\partial \theta} 
\left(sin \theta cos \theta \frac{\partial Y}{\partial \theta}
\right)
- \left(\frac{k^2}{(cos \theta)^2}+ \frac{m^2}{(sin \theta)^2} \right)
Y = \lambda Y.
\end{equation}

Going over to the tortoise coordinate $d r^{*} =d r/f(r)$ and to the new
wave function $\Psi = R(r) r ^{3/2}$,  the equation (13) can be
reduced, after some algebra, to the wave-like form

\be\label{Wave-like-equation}%
\left(\frac{d^2}{dr^{*2}} + \om^2 + 4 j \omega (k + m) - V(r^*)\right)\Psi = 0. \ee%

The effective potential has the form
\begin{equation}
V(r)=f(r)\left(\frac{3}{4 r^{2}}f(r)+ \frac{3}{2r} f'(r)+ \frac{(2 l + k + m) (2 l + k + m + 2)}{r^{2}}\right).
\end{equation}
Here $l$, $k$, and $m$ run over the values $0, 1, 2, ...$
The tortoise coordinate $r^{*}$ is defined on the interval $(- \infty,  +
\infty)$ in such a way, that the spatial infinity
$r=+\infty $ corresponds to $r^{*} =\infty$, while the event horizon
corresponds to   $r^{*} =-\infty$. 
The above effective potential is positively defined  and has the form of 
the potential barrier which approaches constant values at both spatial
infinity and event horizon. In fact, the potential $V(r)$ coincides
with that for (4+1)-dimensional  Schwarzschild  black hole when
taking the multipole number $L$ to be $2 l + k + m$, yet, the spectrum
is different, due to the term  $4 j \omega (k + m)$ in (5),
which depend  not only on the final value $L$, but also on   
terms $l$, $k$, $m$.

\section{Quasinormal modes of the Schwarzschild-G\"{o}del black hole.}

If choosing a positive sign for the real part of
$\omega$  ($\omega = Re \omega - i Im \omega$), QNMs satisfy the following boundary conditions
\be\label{bounds} \P(r^*) \sim C_\pm \exp(\pm\imo\om r^*), \qquad
r\longrightarrow \pm\infty,
\ee%
corresponding to purely in-going waves at the event horizon and purely out-going waves at infinity.

In order to find quasinormal frequencies of the black hole with an effective
potential in the form of the potential barrier (16),  we use the WKB approach. 
The WKB approach  was used for calculations of quasinormal modes 
in the first order beyond eikonal approximation by Schutz and Will 
\cite{schutz-will}, extended by Iyer and Will to the third 
order \cite{IyerWill}, and recently extended to the sixth 
WKB order \cite{KonoplyaWKB6}. The WKB approach up to the 6th WKB
order has been used used recently in a series of papers
\cite{cardoso-acustic}, \cite{berti-kok}, \cite{konoplya03-3}, where QN
frequencies of different black holes were considered, and,
the comparison with accurate numerical values showed very good
agreement. The accuracy of the WKB results is the better, the larger
the multipole number $L$ and the less the overtone number $n$. In fact 
for $n$ larger then $L$ the WKB formula cannot be applied.  

From here we shall use the units such that $2 M =1$.

The WKB formula has the form \cite{KonoplyaWKB6}:
 
\be\label{WKB}
\imo\frac{\om^2 - V_0}{\sqrt{-2V_0^{\prime\prime}}} - \L_2 - \L_3 - \L_4 -
\L_5 - \L_6 = n + \frac{1}{2},
\ee
where $V_0$ is the height and $V_0^{\prime\prime}$ is the second
derivative with respect to the tortoise coordinate of the
potential at the maximum. $\L_2$ and $\L_3$ can be found in
\cite{IyerWill}, $\L_4$, $\L_5$ and $\L_6$ are presented in
\cite{KonoplyaWKB6}; the corrections depend on the value of the potential
and higher derivatives of it at the maximum.

The 6th order WKB values of the quasinormal frequencies are shown on  
Figures 1-4 and Table I. 
From Fig. 1 one can see that the real part of $\omega$ is decreasing
with growing of $j$, being roughly proportional to $j$, for a fixed
black hole mass and fixed values of $l$, $m$, $k$.
This can be easily explained in the following way: 
from the wave equation (15) one can learn that if one discards small values of
order $O(j^2)$,  then $\omega^{2} + 4 j (m+k) \omega = \omega_{0}^2$, 
where $\omega_{0}$ is the Schwarzschild value of $\omega$
under some fixed $M$, $l$, $m$, $k$, $n$. Furthermore this can be
represented as $(\omega + 2 j (k+m))^{2} - 8 j^2 (k+m)^2=\omega_{0}^2$,
i.e. the real part of $\omega$ is roughly increased by  $2 j
(k+m)$, while the change in imaginary part comes from the term $8 j^2 (k+m)^2$. 
More accurately, Fig. 2 shows that the imaginary part is decreasing when
$j$ is increasing. Therefore, the influence of rotating cosmological
background, represented by parameter $j$, gives rise to decreasing of
the oscillation frequency and of the damping rate. Thus, in the
rotating Universe the QN modes damp more slowly, but, because of the 
considerable falling down of $Re \omega$ and slight falling down 
of the  $Im \omega$, the resulting quality factor $Re \omega/2 Im \omega$
is decreasing and, thereby, the black hole in a rotating Universe
is a worse oscillator than in a non-rotating one.

In Table 1. we put the low overtones of the QN spectrum for different
values of $k$, $m$, $l$, and $j$. From that table we can learn that
the higher any of the values $k$, $l$, or $m$ and the lower the
overtone $n$, the better the accuracy of the WKB formula, and, as a
result, the less is the difference between the 6th and 3th order WKB data.
An intrinsic fact for  any black hole quasinormal spectrum, when the
overtone number grows $Re \omega$ falls down, while the damping
rate grows. We cannot judge what will happen with asymptotically high 
overtones ($n \rightarrow \infty$), since we analyze here only 
an approximate solution.

In the eikonal (high frequency) approximation, we can use the first
order WKB formula for finding the lower overtones. Thus, for large
$l$, and thereby for large $L= 2 l + k + m$, in units  $2 (2 M)^{-1}$ we obtain:
\begin{equation}  
\omega = \frac{L+1}{2} + 2 j (k + m) - i \frac{2 n +1}{2 \sqrt{2}}.
\end{equation}
Note that there are two limitations on this formula.
First, when $k$ or $m$ is also large, the general relation 
\begin{equation}  
\omega^2 - 4 j (k+m) \omega = \left(\frac{L+1}{2} - i \frac{2 n +1}{2 \sqrt{2}}\right)^2
\end{equation} 
holds. Moreover, we should be careful when interpreting this formula at
asymptotically large $L$, since it uses an approximate metric
and $j^{2}$ corrections in metric may produce different asymptotic values.
Yet we expect it should be correct for moderately large values of $L$.
Note that the above formulas are accurate enough even for not very large
values of $L$, for instance, for $L=4$ the relative error is about
several percents.

The massless scalar QNMs analysis can easily be extended to the massive
case, in which one has the same wave-like equation but with the
effective potential 
\begin{equation}
V(r)=f(r)\left(\frac{3}{4 r^{2}}f(r)+ \frac{3}{2r} f'(r)+ \frac{(2 l +
k + m) (2 l + k + m + 2)}{r^{2}} +\mu^2 \right).
\end{equation} 
Yet we can use the above WKB formula only for small values of the 
field mass $\mu$, since for large $\mu$ the effective potential has
three turning points. The 6th order WKB frequencies are presented in
Fig.3 and 4  as functions of $\mu$. Thus, we see that the larger the field
mass, the larger is  the  real part of $\omega$,  and  the smaller the
imaginary part. 
In other words, the ``massive'' QN modes decay more slowly and have 
greater real frequency of oscillation. It is known that at  asymptotically high
overtones the mass of the field does not affect the QN modes \cite{SMF}. 
Note that all the above features were observed for massive scalar
field of the Schwartzshild \cite{SMF} and Reissner-Nordstrom black
holes \cite{RMF}, \cite{Xue}.Since these features were found also for massive
Dirac field (see \cite{Karlucio} and references
therein), it is possible that they are  generic for massive fields of arbitrary spin.

\vspace{4mm}
      
\begin{minipage}[c]{.45\textwidth}
TABLE I: WKB values for QNMs at fixed $j=1/8$. $2 M=1$.
\begin{longtable}{|c|c|c|c|c|c|}
 $k$& $m$& $l$  & $n$  & 3th WKB order & 6th WKB order \\ 
\hline 
 $0$& $0$& $0$& $0$ & $ 0.49123 - 0.41100 i $ & $ 0.54633 - 0.36087 i $\\ 
 $1$& $0$& $0$& $0$ & $ 0.78080 - 0.35384 i $ & $ 0.79151 - 0.35575 i $ \\ 
 $1$& $0$& $0$& $1$ & $ 0.59794 - 1.15113 i $ & $ 0.61846 - 1.14902 i $ \\ 
 $0$& $0$& $1$& $0$ & $ 1.50707 - 0.35787 i $ & $ 1.51050 - 0.35770 i $  \\
 $0$& $0$& $1$& $1$ & $ 1.38524 - 1.10754 i $ & $ 1.39249 - 1.10537 i $ \\ 
 $0$& $0$& $1$& $2$ & $ 1.19442 - 1.90690 i $ & $ 1.18639 - 1.94829 i $ \\ 
 $1$& $0$& $1$& $0$ & $ 1.77155 - 0.35330 i $ & $ 1.77293 - 0.35317 i $ \\ 
 $1$& $0$& $1$& $1$ & $ 1.67618 - 1.07942 i $ & $ 1.67927 - 1.07838 i $ \\ 
 $1$& $0$& $1$& $2$ & $ 1.51426 - 1.84424 i $ & $ 1.50511 - 1.86038 i $\\ 
 $1$& $0$& $1$& $3$ & $ 1.30766 - 2.63920 i $ & $ 1.27639 - 2.73256 i $\\ 
 $1$& $1$& $1$& $0$ & $ 2.05408 - 0.34832 i $ & $ 2.05475 - 0.34826 i $\\ 
 $1$& $1$& $1$& $1$ & $ 1.97396 - 1.05821 i $ & $ 1.97552 - 1.05774 i $\\ 
 $1$& $1$& $1$& $2$ & $ 1.83110 - 1.79890 i $ & $ 1.82449 - 1.80621 i $\\ 
 $1$& $1$& $1$& $3$ & $ 1.64511 - 2.57051 i $ & $ 1.61726 - 2.61762 i $\\ 
 $1$& $1$& $1$& $4$ & $ 1.42547 - 3.36434 i $ & $ 1.37379 - 3.51096 i $\\ 
 \end{longtable}
\end{minipage} 

\vspace{4mm}

\section{Discussions}

We have investigated the decay of (generally speaking, massive) scalar
field around a Schwartzshild black hole immersed in a rotating 
cosmological background. In the limit of the small cosmological 
parameter $j$, the QNMs, which govern the decay of the scalar field
at late times, have been found. It was found that the cosmological 
rotation gives rise the decreasing of the real frequencies of
oscillations (proportional to the cosmological parameter) 
and of damping rates. The quality factor of the black hole
as an oscillator is smaller in the presence of cosmological rotation.
The massive scalar field damps more slowly and have greater oscillation
frequency. All found modes are damping what supports the stability of
the Schwartzshild-G\"{o}del space-time against scalar field perturbations.
Yet, within the approximate solution we analyzed, one cannot judge
about stability eventually. Note also that the stability of the 
metric as such is determined by the gravitational perturbations,
although the scalar field perturbations may coincide with tensor type
gravitational perturbations \cite{ishibashi} which are decisive in gravitational
stability \cite{gibbons}. The present analysis can also be extended to
the case of scalar field interacting electromagnetically with the
charge of the black hole \cite{charged_scalar}, i.e. to the case of the decay of charged
scalar field around a Reissner-Nordstrem-G\"{o}del black hole.


\begin{figure}
\begin{center}
\includegraphics{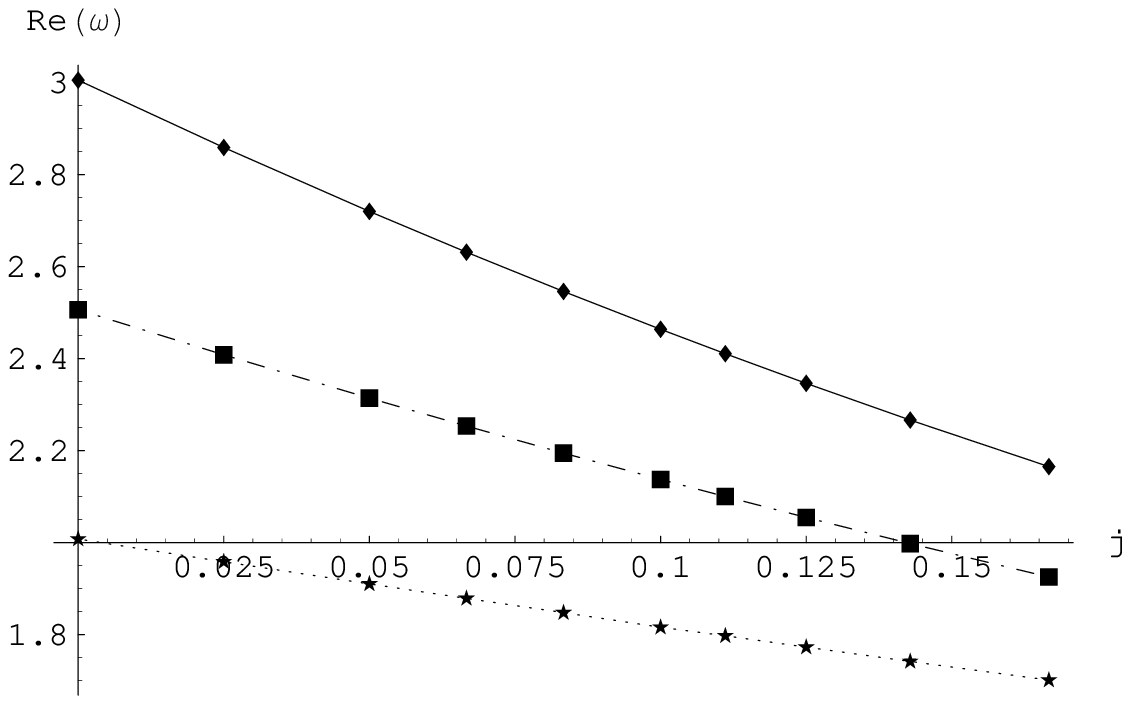}
\caption{Real part of $\omega$ as a function of the G\"{o}del background scale
parameter $j$ (diamond $k=2$, $m=1$, $l=1$), (box $k=1$, $m=1$,
$l=1$), (star $k=0$, $m=1$, $l=1$).}
\label{1}
\end{center}
\end{figure}

\begin{figure}
\begin{center}
\includegraphics{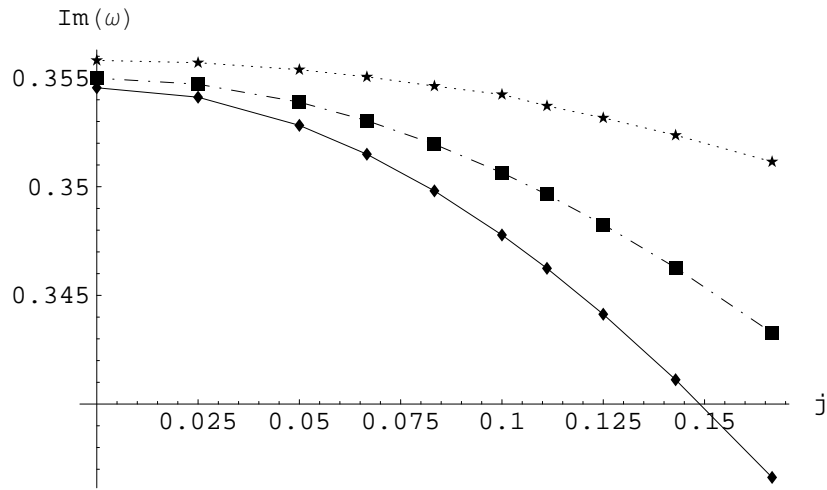}
\caption{Imaginary part of $\omega$ as a function of the G\"{o}del background scale
parameter $j$ (star $k=2$, $m=1$, $l=1$), (box $k=1$, $m=1$,
$l=1$), (diamond $k=0$, $m=1$, $l=1$).}
\label{2}
\end{center}
\end{figure}

\begin{figure}
\begin{center}
\includegraphics{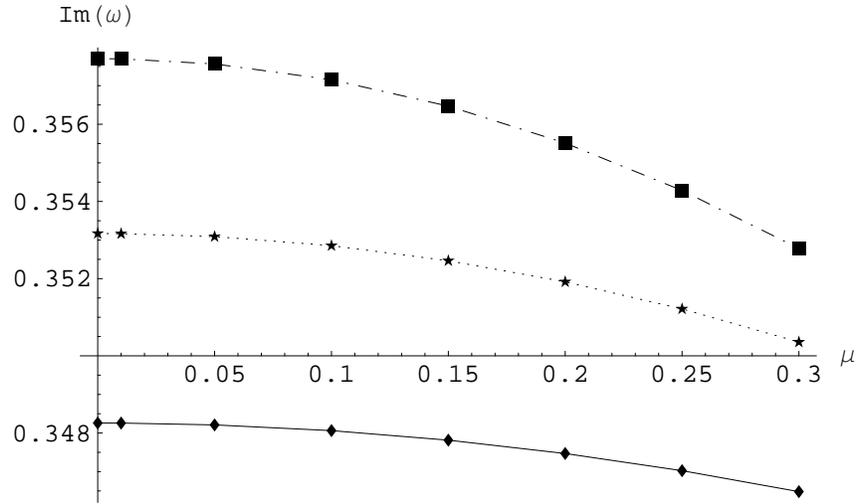}
\caption{Imaginary part of $\omega$ as a function of the mass $\mu$ 
for $l=1$, $k=1$, $m=1$ (box), $l=1$, $k=0$, $m=1$ (star), $l=1$,
$k=0$, $m=0$ (diamond); $j=1/8$.}
\label{3}
\end{center}
\end{figure}

\begin{figure}
\begin{center}
\includegraphics{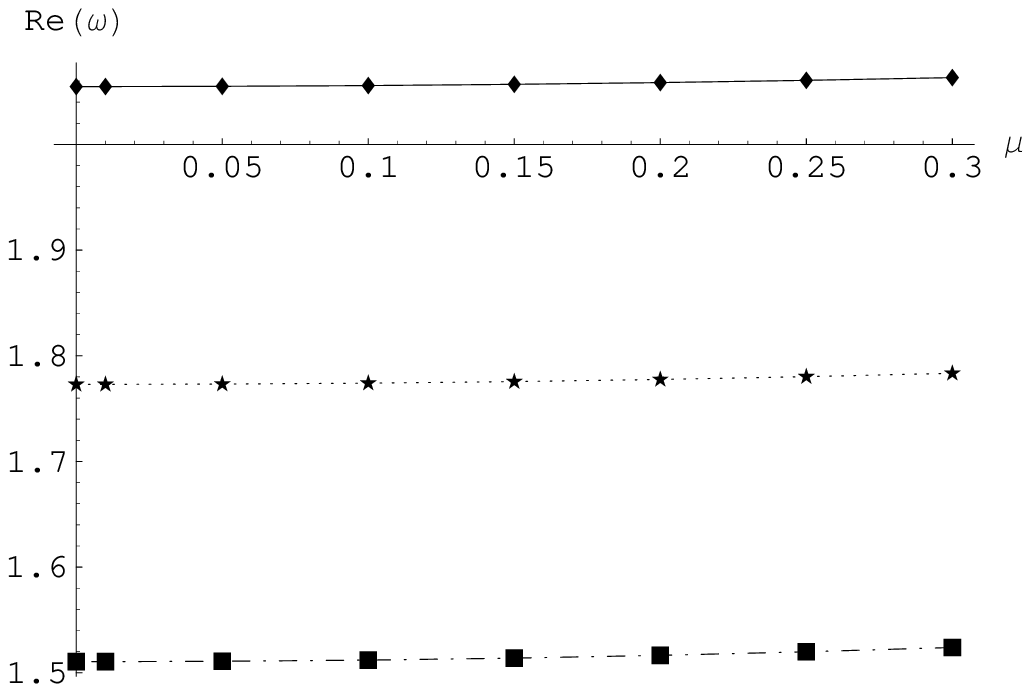}
\caption{Real part of $\omega$ as a function of the mass 
$\mu$ for $l=1$, $k=1$, $m=1$ (box), $l=1$, $k=0$, $m=1$ (star), $l=1$, $k=0$, $m=0$ (diamond); $j=1/8$.}
\label{4}
\end{center}
\end{figure}

\end{document}